\documentclass{article}

\usepackage{arxiv}

\usepackage[utf8]{inputenc} 
\usepackage[T1]{fontenc}    
\usepackage{hyperref}       
\usepackage{url}            
\usepackage{booktabs}       
\usepackage{amsmath}        
\usepackage{amsfonts}       
\usepackage{amssymb}
\usepackage{nicefrac}       
\usepackage{microtype}      
\usepackage{graphicx}
\usepackage[numbers,sort&compress]{natbib}
\usepackage{doi}
\usepackage{float}          
\usepackage{longtable}      
\usepackage{booktabs}
\usepackage{siunitx}
\usepackage{subfig}         
\usepackage{tikz}
\usetikzlibrary{arrows.meta, positioning}
\usepackage{lineno}         

\newcommand{\authorcontributions}[1]{\paragraph{Author Contributions.} #1}
\newcommand{\funding}[1]{\paragraph{Funding.} #1}
\newcommand{\institutionalreview}[1]{\paragraph{Institutional Review Board Statement.} #1}
\newcommand{\informedconsent}[1]{\paragraph{Informed Consent Statement.} #1}
\newcommand{\dataavailability}[1]{\paragraph{Data Availability Statement.} #1}
\newcommand{\acknowledgments}[1]{\paragraph{Acknowledgments.} #1}
\newcommand{\conflictsofinterest}[1]{\paragraph{Conflicts of Interest.} #1}
\newcommand{\abbreviations}[2]{\paragraph{#1.} #2}
\newcommand{\reftitle}[1]{}
\newenvironment{adjustwidth}[2]{}{}

\title{A Bayesian Time-Varying SEIARD Model for State-Level COVID-19 Transmission and Mortality in the United States}

\author{ \href{https://orcid.org/0000-0002-6218-615X}{\includegraphics[scale=0.1]{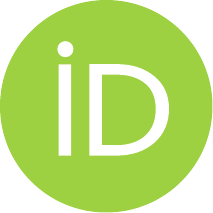}\hspace{1mm}Paromita Banerjee} \\
	Department of Mathematics, Computer Science, and Data Science\\
	John Carroll University\\
	University Heights, OH 44118, USA \\
	\texttt{pbanerjee@jcu.edu} \\
}

\renewcommand{\shorttitle}{Bayesian Time-Varying SEIARD Model for COVID-19}

\hypersetup{
pdftitle={A Bayesian Time-Varying SEIARD Model for State-Level COVID-19 Transmission and Mortality in the United States},
pdfsubject={q-bio.PE, stat.AP},
pdfauthor={Paromita Banerjee},
pdfkeywords={COVID-19, Compartmental Model, Bayesian inference, Markov chain Monte Carlo (MCMC), Robust clustering},
}

\begin{document}
\maketitle

\begin{abstract}
We conduct a retrospective analysis of COVID-19 transmission dynamics across U.S. states using a modified population-based Susceptible-Exposed-Infectious-Asymptomatic-Recovered-Deceased (SEIARD) compartmental model. The proposed framework introduces time-varying transmission, reporting, and mortality rates to capture temporal variations in public behavior and policy interventions during the pandemic. In particular, the transmission rate is modeled as a function of population mobility (derived from Google Mobility Reports), with a residual time-decay term capturing the net effect of unobserved factors such as behavioral adaptation and control measures, while reporting is linked to nationwide testing strategies. We employ a Bayesian approach to integrate multiple data sources and quantify uncertainties in model parameters. The model explicitly distinguishes between symptomatic and asymptomatic infectious individuals and links the latent epidemic states to observable quantities, including reported cases and deaths, through a dynamic reporting function. This retrospective modeling framework provides insights into state-level epidemic trajectories and supports data-driven decision-making for optimal allocation of healthcare resources and evaluation of public health interventions during future pandemics. We further apply a clustering analysis to the posterior parameter estimates to identify groups of U.S. states exhibiting similar epidemiological characteristics, revealing substantial regional heterogeneity in transmission intensity, reproduction dynamics, and mortality burden.
\end{abstract}


\section{Introduction}

The global outbreak of coronavirus disease 2019 (COVID-19), caused by the severe acute respiratory
syndrome coronavirus 2 (SARS-CoV-2), has posed unprecedented challenges to public health systems,
economies, and societies worldwide since its emergence in late 2019 \citep{who2020}. Mathematical
and statistical models have played a central role in understanding the transmission dynamics of the
disease, forecasting epidemic trajectories, and assessing the effectiveness of control strategies
such as social distancing, mask mandates, vaccination, and testing \citep{fang2020,giordano2020,li2020}.

Among these, compartmental models, such as the classical
Susceptible-Exposed-Infectious-Recovered (SEIR) framework, have been widely used to describe
the temporal evolution of infectious diseases \citep{kermack1927}. However, the complexity of
COVID-19 epidemiology, including asymptomatic transmission, heterogeneous reporting, and
time-varying contact behaviors, has motivated the development of extended models that more
accurately capture real-world dynamics \citep{arenas2020,hao2020}. In particular, the inclusion
of an asymptomatic compartment and disease-induced mortality has led to the SEIARD structure,
which provides a more realistic representation of SARS-CoV-2 transmission and outcomes
\citep{gatto2020,pelinovsky2020}.

Similar compartmental models for U.S.\ states have been considered by several research groups
\citep{chiu2020state,khan2020predictive,yarsky2021genetic}. These studies largely focus on the
initial phase of disease progression without accounting for a time-dependent
transmission rate. The present work advances on these studies by incorporating a time-dependent
transmission rate that accounts for changes in public interaction and activities during the pandemic
(estimated from Google Mobility data), together with a residual decay component representing the
combined, unobserved effects of behavioral adaptation, such as mask-wearing and self-quarantine,
and other control measures not directly captured by mobility. Furthermore,
prior research has focused on calibrating deterministic SEIR models using optimization methods, which
yields point estimates that lack quantification of parameter uncertainty. Here we adopt a Bayesian
approach in which uncertainty in model parameters is naturally expressed through posterior probability
distributions. A further contribution of this work is the use of a Bayesian model to
integrate multiple data sources, including infection rates, death rates, mobility, and testing rates,
with efficient adaptive Markov chain Monte Carlo (MCMC) methods employed for sampling-based inference.

Transmission rates during the pandemic were not static but fluctuated in response to policy
interventions, behavioral adaptations, and mobility changes. Empirical studies using Google Mobility
Reports have demonstrated strong correlations between mobility reductions and decreases in COVID-19
transmission \citep{kraemer2020,nourein2022}. Incorporating such data into transmission models
enables dynamic parameterization of infection rates, reflecting temporal variations in public
interaction levels and adherence to mitigation measures. Additionally, testing intensity influences
both case detection and the effective reporting rate, serving as a proxy for the surveillance
capacity of public health systems \citep{kuhn2021}. Modeling these factors jointly allows for a
more comprehensive understanding of the epidemiological processes underlying observed case and
mortality trends.

In this study, we propose a time-varying SEIARD model that incorporates mobility-driven
transmission with a residual behavioral-decay component, along with testing-linked reporting,
and we employ a Bayesian framework to quantify parameter uncertainty across U.S.\ states. To
characterize regional structure in the epidemic, we further apply a clustering analysis to the
posterior parameter estimates, grouping states with similar transmission, reproduction, and
mortality profiles to reveal regional epidemiological patterns.

The remainder of this paper is organized as follows. Section~\ref{sec:seiard} presents the SEIARD
model formulation. Section~\ref{sec:extended} introduces the extended model with mobility and testing
dependence. Section~\ref{sec:bayesian} describes the parameter estimation
and Bayesian inference framework, with the adaptive MCMC algorithm detailed in
Section~\ref{sec:adaptivemcmc}. Section~\ref{sec:results} reports results for U.S.\ states,
including the state-level clustering analysis. Section~\ref{sec:discussion} discusses the findings
and limitations, and
Section~\ref{sec:conclusions} concludes.

\section{SEIARD Model Formulation}
\label{sec:seiard}

We employ a deterministic compartmental model of COVID-19 transmission of type SEIARD
(Susceptible-Exposed-Infectious-Asymptomatic-Recovered-Deceased), extended to include
time-varying transmission, reporting, and mortality rates. The total population is assumed constant
and denoted by $N$.

\subsection{Model Compartments}

The model partitions the population into six compartments. Here $S(t)$ denotes susceptible individuals; $E(t)$ denotes exposed individuals, who are infected but not yet infectious; $I(t)$ denotes symptomatic infectious individuals, who are subject to reporting and disease-induced death; $A(t)$ denotes asymptomatic infectious individuals; $R(t)$ denotes recovered individuals, who are immune; and $D(t)$ denotes cumulative disease-induced deaths. In addition, we track two observable cumulative quantities:
$
C_{\mathrm{rep}}(t) = \text{cumulative reported cases}$, and
$C_{\mathrm{death}}(t) = \text{cumulative reported deaths}.
$

\subsection{Model Parameters}

The principal parameters of the model are as follows: $\beta(t)$ denotes the time-varying
transmission rate; $\sigma$ denotes the rate of progression from exposed to infectious
(the inverse of the mean latent period); $\theta$ denotes the fraction of infections that become
symptomatic; $\gamma_I$ and $\gamma_A$ denote the recovery rates for symptomatic and asymptomatic
infectious individuals, respectively; $\mu(t)$ denotes the time-varying disease-induced death rate
for symptomatic infectious individuals; and $p(t)$ denotes the time-varying fraction of new
symptomatic infections that are reported.

\subsection{Model Equations}

Let the \emph{force of infection} be
\[
\lambda(t) = \beta(t)\,\frac{I(t) + A(t)}{N}.
\]
The system of ordinary differential equations (ODEs) governing the model is
\begin{linenomath}
\begin{align}
\frac{dS}{dt} &= -\lambda(t)\,S, \label{eq:S}\\[4pt]
\frac{dE}{dt} &= \lambda(t)\,S - \sigma\,E, \label{eq:E}\\[4pt]
\frac{dI}{dt} &= \theta\,\sigma\,E - \bigl(\gamma_I + \mu(t)\bigr)\,I, \label{eq:I}\\[4pt]
\frac{dA}{dt} &= (1-\theta)\,\sigma\,E - \gamma_A\,A, \label{eq:A}\\[4pt]
\frac{dR}{dt} &= \gamma_I\,I + \gamma_A\,A, \label{eq:R}\\[4pt]
\frac{dD}{dt} &= \mu(t)\,I. \label{eq:D}
\end{align}
\end{linenomath}

Reported (observable) flows are given by
\begin{linenomath}
\begin{align}
\frac{dC_{\mathrm{rep}}}{dt}   &= p(t)\,\theta\,\sigma\,E, \label{eq:Crep}\\[4pt]
\frac{dC_{\mathrm{death}}}{dt} &= \mu(t)\,I. \label{eq:Cdeath}
\end{align}
\end{linenomath}

Initial conditions at $t=0$ are specified as
\[
S(0)=S_0,\quad E(0)=E_0,\quad I(0)=I_0,\quad A(0)=A_0,\quad R(0)=R_0,\quad D(0)=D_0,
\]
with $S_0+E_0+I_0+A_0+R_0+D_0 = N$ and $C_{\mathrm{rep}}(0)=C_{\mathrm{death}}(0)=0$.

\subsection{Reproduction Number}

The instantaneous basic reproduction number $R_0(t)$, defined as the expected number of
secondary infections produced by a typical infectious individual in a fully susceptible population, is
\begin{linenomath}
\begin{equation}
R_0(t) \;=\;
\beta(t)\!\left(\frac{\theta}{\gamma_I+\mu(t)} + \frac{1-\theta}{\gamma_A}\right).
\end{equation}
\end{linenomath}
Accounting for depletion of susceptibles, the effective reproduction number is
\begin{linenomath}
\begin{equation}
R_e(t) \;=\; \frac{S(t)}{N}\,R_0(t).
\end{equation}
\end{linenomath}

\subsection{Model Assumptions}

The model rests on the following simplifying assumptions. The population is assumed to mix
homogeneously, with no age or spatial structure. Natural births and non-disease deaths are neglected,
so that the total population $N$ remains constant. Disease-induced mortality is assumed to occur
exclusively within the symptomatic infectious class $I(t)$. The reporting fraction $p(t)$ is applied
to the flow of newly symptomatic infections from $E$ to $I$, so that only a fraction of incident
symptomatic cases enters the reported case count. Finally, reporting of deaths is assumed complete,
so that $C_{\mathrm{death}}(t)$ faithfully reflects the true cumulative mortality.

\subsection{Model Extensions}

The baseline SEIARD structure presented here admits several natural generalizations. Hospitalization
compartments can be incorporated to capture capacity-dependent mortality and to link model outputs to
hospital admission data, which are often more reliably reported than raw case counts. Explicit
reporting delay distributions, modeled, for example, through convolution kernels or delay
differential equations, would allow the framework to account for the lag between infection onset and
case confirmation. The time-varying parameters $\beta(t)$, $\mu(t)$, and $p(t)$ can be further
linked to additional observed covariates such as policy stringency indices, vaccination coverage, or
meteorological factors, following the covariate regression structure introduced in
Section~\ref{sec:extended}. Finally, stochastic formulations, using Poisson or negative binomial
observation models, or continuous-time Markov chain dynamics, can replace the deterministic ODE
layer to better characterize variability in low-incidence settings or at the onset of an outbreak.

\section{Extended SEIARD Model with Mobility and Testing Dependence}
\label{sec:extended}

To capture behavioral and policy effects on transmission and case detection, we extend the SEIARD
framework to include the influence of population mobility and diagnostic testing rates. Specifically,
the transmission rate $\beta(t)$ and reporting fraction $p(t)$ are expressed as functions of
time-dependent covariates: relative mobility $M(t)$ and testing rate $T(t)$.

\subsection{Model Structure}

The extended model retains the six-compartment SEIARD structure defined in Section~\ref{sec:seiard},
with state variables $S(t)$, $E(t)$, $I(t)$, $A(t)$, $R(t)$, and $D(t)$ governed by the ODE
system in Equations~(\ref{eq:S}) to (\ref{eq:D}). The total population $N$ is held constant, and the
two cumulative observable quantities $C_{\mathrm{rep}}(t)$ and $C_{\mathrm{death}}(t)$ are tracked
as before. The key extension introduced here is to express the transmission rate $\beta(t)$ and the
reporting fraction $p(t)$ as explicit functions of time-varying observable covariates, as detailed
in the subsections below.

\subsection{Time-Varying Parameters Linked to Covariates}

The transmission rate $\beta(t)$ and reporting fraction $p(t)$ are modeled as smooth functions of
normalized mobility and testing rate indices, $M(t)$ and $T(t)$, respectively.

\paragraph{Transmission rate.}
Mobility affects the average contact rate between individuals. Let $M(t)$ denote the relative
mobility at time $t$ (normalized so that $M=1$ corresponds to pre-pandemic baseline mobility). We
define:
\begin{linenomath}
\begin{equation}
\beta(t) = \beta_0 \,[1 + \alpha_M \,(M(t) - 1)] \, e^{-\rho_\beta\,t},
\label{eq:beta_mobility}
\end{equation}
\end{linenomath}
where $\beta_0$ is the baseline transmission rate, $\alpha_M$ quantifies the sensitivity of
transmission to changes in mobility, and $\rho_\beta$ represents gradual behavioral adaptation or
residual control measures independent of mobility (e.g., mask use). A positive $\alpha_M$ implies
that transmission increases with mobility.

\paragraph{Reporting fraction.}
Testing availability influences the proportion of symptomatic cases detected and reported. Let
$T(t)$ denote the normalized daily testing rate (e.g., tests per thousand individuals). We define:
\begin{linenomath}
\begin{equation}
p(t) = p_{\min} + \frac{p_{\max} - p_{\min}}{1 + \exp[-\alpha_T (T(t) - T_0)]},
\label{eq:report_testing}
\end{equation}
\end{linenomath}
where $p_{\min}$ and $p_{\max}$ are the lower and upper bounds of reporting probability, $\alpha_T$
is the responsiveness of reporting to testing intensity, and $T_0$ represents the inflection point
(testing rate at which reporting probability reaches its midpoint). This logistic form ensures
$p(t)\in[p_{\min},p_{\max}]$.

\paragraph{Mortality rate.}
Clinical outcomes improve with experience and healthcare adaptation; we allow the death rate among
symptomatic infectious individuals to decrease exponentially:
\begin{linenomath}
\begin{equation}
\mu(t) = \mu_0 \, e^{-\rho_\mu\,t},
\label{eq:mu_decay}
\end{equation}
\end{linenomath}
where $\mu_0$ is the initial daily death rate and $\rho_\mu$ is the rate of clinical improvement.

\subsection{Model Equations}

Let the \emph{force of infection} be
\[
\lambda(t) = \beta(t)\,\frac{I(t) + A(t)}{N}.
\]
The SEIARD dynamics are given by Equations~(\ref{eq:S}) to (\ref{eq:D}), with reported quantities
evolving according to:
\begin{linenomath}
\begin{align}
\frac{dC_{\mathrm{rep}}}{dt}   &= p(t)\,\theta\,\sigma\,E, \label{eq:Crep2}\\[4pt]
\frac{dC_{\mathrm{death}}}{dt} &= \mu(t)\,I. \label{eq:Cdeath2}
\end{align}
\end{linenomath}

\subsection{Reproduction Number}

The instantaneous basic reproduction number $R_0(t)$ for the above model is given by
\begin{linenomath}
\begin{equation}
R_0(t) =
\beta(t)\!\left(\frac{\theta}{\gamma_I+\mu(t)} + \frac{1-\theta}{\gamma_A}\right),
\end{equation}
\end{linenomath}
and the effective reproduction number, accounting for the susceptible fraction, is given by
\begin{linenomath}
\begin{equation}
R_e(t) = \frac{S(t)}{N}\,R_0(t).
\end{equation}
\end{linenomath}

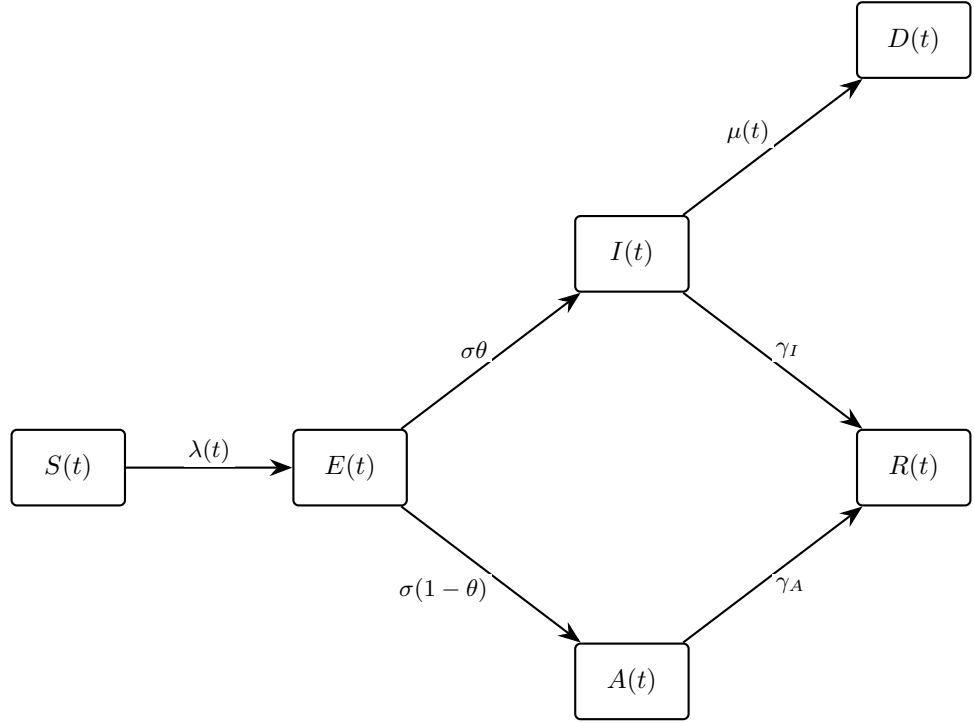
\begin{figure}[H]
    \centering
    \begin{tikzpicture}[
        >={Stealth[length=3mm]},
        node distance=18mm and 22mm,
        box/.style={
            draw, rounded corners=2pt, thick,
            minimum width=15mm, minimum height=10mm,
            font=\normalsize
        },
        flow/.style={->, thick},
        lbl/.style={font=\small, inner sep=1.5pt, fill=white}
    ]
        \node[box] (S) {$S(t)$};
        \node[box, right=of S] (E) {$E(t)$};
        \node[box, above right=of E] (I) {$I(t)$};
        \node[box, below right=of E] (A) {$A(t)$};
        \node[box, above right=of I] (D) {$D(t)$};
        \node[box, below right=of I] (R) {$R(t)$};

        \draw[flow] (S) -- node[lbl, above] {$\lambda(t)$} (E);
        \draw[flow] (E) -- node[lbl, above left=-1pt] {$\sigma\theta$} (I);
        \draw[flow] (E) -- node[lbl, below left=-1pt] {$\sigma(1-\theta)$} (A);
        \draw[flow] (I) -- node[lbl, above left=-1pt] {$\mu(t)$} (D);
        \draw[flow] (I) -- node[lbl, above right=-1pt] {$\gamma_I$} (R);
        \draw[flow] (A) -- node[lbl, below right=-1pt] {$\gamma_A$} (R);
    \end{tikzpicture}
    \caption{Schematic diagram of the SEIARD model.\label{fig:SchDiag}}
\end{figure}

A schematic diagram of the modified SEIARD model is given in Figure~\ref{fig:SchDiag}.

\subsection{Interpretation}

In this formulation, mobility directly modulates transmission intensity, while the testing rate
influences the visibility of the epidemic via $p(t)$. High mobility increases the contact rate
and thus $R_e(t)$, whereas increased testing raises the fraction of infections detected
without affecting true transmission dynamics. Together, $(M(t),T(t))$ serve as observable covariates
linking social behavior and surveillance capacity to epidemic trajectories.



\section{Parameter Estimation and Uncertainty Quantification using the Bayesian Method}
\label{sec:bayesian}

We cast the deterministic SEIARD ODE model into a Bayesian state-space formulation by coupling
the ODE dynamics to an observation model for reported cases and deaths.

\paragraph{Latent deterministic dynamics:}
Let $X(t) = \bigl(S(t),E(t),I(t),A(t),R(t),D(t)\bigr)^\top$ denote the ODE state at continuous
time $t$ governed by
\[
\frac{dX}{dt} = f\bigl(X(t); \boldsymbol{\phi}, \beta(\cdot), \mu(\cdot), p(\cdot)\bigr),
\]
with initial condition $X(0)=X_0$. The function $f(\cdot)$ is the SEIARD right-hand side and
depends on epidemiological parameters collected in $\boldsymbol{\phi}$ (e.g.,
$\sigma,\gamma_I,\gamma_A,\theta$) and on parameterized time-varying functions for transmissibility
$\beta(t)$, mortality $\mu(t)$, and reporting $p(t)$. 

For inference we represent the time-varying
functions with low-dimensional parameterizations. Reporting is given the four-parameter logistic
form defined in Equation~(\ref{eq:report_testing}),
$p(t)=p_{\min}+\dfrac{p_{\max}-p_{\min}}{1+\exp[-\alpha_T(T(t)-T_0)]}$, where $T(t)$ is the
normalized daily testing rate and $(p_{\min},p_{\max},\alpha_T,T_0)$ are estimated; $\beta(t)$ is
represented through the mobility-linked functional form of Equation~(\ref{eq:beta_mobility})
(parameters $\beta_0,\alpha_M,\rho_\beta$); and $\mu(t)$ follows the exponential-decay form of
Equation~(\ref{eq:mu_decay}) (parameters $\mu_0,\rho_\mu$).
Collect all unknown parameters in $\boldsymbol{\Theta}=(\boldsymbol{\phi},\boldsymbol{\beta},
\boldsymbol{\mu},\boldsymbol{\psi})$, where $\boldsymbol{\psi}=(p_{\min},p_{\max},\alpha_T,T_0)$ are the reporting parameters. These parameters were calibrated using the Bayesian approach detailed below.

\subsection{Observed data and Likelihood}
Model calibration was performed using daily time series of reported confirmed cases and reported
deaths, denoted by $\{C^{\mathrm{obs}}_t, D^{\mathrm{obs}}_t\}_{t=1}^T$. These data are obtained
from official surveillance systems or public repositories \citep{johnhopkins2020dataset,ourworldindata2021covid}.
Dates are converted to integer time indices ($t=1,2,\ldots,T$), and cumulative series are
differenced to obtain daily incidence for model fitting.
We observe daily reported incidence and deaths at discrete days $t=1,\dots,T$. 

Let
$\widehat{C}_t(\boldsymbol{\Theta})$ and $\widehat{D}_t(\boldsymbol{\Theta})$ be the
model-derived daily reported cases and deaths obtained by integrating the ODE and applying the
reporting function. A flexible observation model is the Negative Binomial:
\begin{linenomath}
\begin{align}
C^{\mathrm{obs}}_t \mid \boldsymbol{\Theta}
  &\sim \operatorname{NegBin}\bigl(\widehat{C}_t(\boldsymbol{\Theta}),\,\kappa_C\bigr),\label{eq:obsC}\\
D^{\mathrm{obs}}_t \mid \boldsymbol{\Theta}
  &\sim \operatorname{NegBin}\bigl(\widehat{D}_t(\boldsymbol{\Theta}),\,\kappa_D\bigr),\label{eq:obsD}
\end{align}
\end{linenomath}
with overdispersion parameters $\kappa_C,\kappa_D>0$. Poisson likelihood arises as a special case when
$\kappa\to\infty$.

\subsection{Priors and Parameterization}
The latent period and recovery rates, $\sigma$, $\gamma_I$, and $\gamma_A$, are held fixed at values
drawn from established estimates in the prior COVID-19 epidemiological literature
\citep{lauer2020incubation,li2020early,he2020temporal,byrne2020inferred} rather than assigned prior distributions:
\[
\sigma = \frac{1}{5.5}, \quad \gamma_I = \frac{1}{7}, \quad \gamma_A = \frac{1}{5.5}.
\]

For the remaining parameters, we choose weakly informative priors that respect parameter
supports and help identifiability. The symptomatic fraction $\theta$ is assigned a
$\operatorname{Beta}(2,2)$ prior, and the log-scale baseline transmission rate $\beta_0$ is given a
$\mathcal{N}(-1.0,\,1.0)$ prior. The mobility sensitivity and decay parameters governing
$\beta(t)$ are assigned $\alpha_M \sim \mathcal{N}(0,\,2^2)$ and
$\rho_\beta \sim \operatorname{Gamma}(0.1,0.1)$, respectively. For the reporting logistic, the bounds
$p_{\min}$ and $p_{\max}$ are given $\operatorname{Beta}(1,9)$ and $\operatorname{Beta}(2,2)$ priors,
the responsiveness parameter $\alpha_T \sim \mathcal{N}(0,\,2^2)$, and the inflection point
$T_0 \sim \operatorname{Beta}(2,2)$. The log-scale baseline mortality rate $\mu_0$ is assigned a
$\mathcal{N}(\log 0.01,\,1.0)$ prior, with its decay parameter
$\rho_\mu \sim \operatorname{Gamma}(0.1,0.1)$. Finally, the overdispersion parameters $\kappa_C$
and $\kappa_D$ are each assigned a $\operatorname{Gamma}(2,0.1)$ prior. Log or logit transforms are
used for positivity or unit-interval constraints; for example, the baseline transmission rate
is represented on the log scale, $\log(\beta_0) \sim \mathcal{N}(-1.0,\,1.0)$, so that the
untransformed parameter $\beta_0$ remains strictly positive.

\subsection{Posterior Distribution}
Let us denote the joint prior distribution of all parameters as  $p(\boldsymbol{\Theta})$, then the full posterior distribution of $\boldsymbol{\Theta}$is given by:
\[ \pi(\boldsymbol{\Theta}) =
p\bigl(\boldsymbol{\Theta}\mid \{C^{\mathrm{obs}}_t,D^{\mathrm{obs}}_t\}_{t=1}^T\bigr)
\;\propto\;
\Bigl[\prod_{t=1}^T f_{\mathrm{NB}}\bigl(C^{\mathrm{obs}}_t\mid\widehat{C}_t(\boldsymbol{\Theta}),\kappa_C\bigr)
f_{\mathrm{NB}}\bigl(D^{\mathrm{obs}}_t\mid\widehat{D}_t(\boldsymbol{\Theta}),\kappa_D\bigr)\Bigr]
\; p(\boldsymbol{\Theta}),
\]
where $f_{\mathrm{NB}}$ denotes the Negative Binomial pmf.

\section{Adaptive Metropolis MCMC for Posterior Sampling}
\label{sec:adaptivemcmc}

We perform Bayesian parameter estimation for the SEIARD model using an adaptive Metropolis Markov
chain Monte Carlo algorithm. Adaptive MCMC automatically tunes the proposal during sampling,
removing (or reducing) the need for manual tuning while preserving ergodicity under mild conditions
\citep{haario2001adaptive,roberts2007coupling,roberts2009examples}.

\subsection{Target Posterior}

Let $\boldsymbol{\Theta}$ denote the vector of model parameters (epidemiological parameters,
time-varying function coefficients, reporting parameters, and observation overdispersion). The
target density is the posterior
\[
\pi(\boldsymbol{\Theta}) \propto \mathcal{L}(\text{data}\mid\boldsymbol{\Theta}) \, p(\boldsymbol{\Theta}),
\]
where $\mathcal{L}$ is the likelihood obtained by numerically integrating the SEIARD ODE and
applying the Negative Binomial observation model.

\subsection{Adaptive Metropolis Algorithm}

We use the Adaptive Metropolis (AM) algorithm of \citet{haario2001adaptive} with a multivariate
Gaussian random-walk proposal whose covariance is adapted online using the empirical covariance of
the chain. Let $d$ be the dimension of $\boldsymbol{\Theta}$ and denote the chain after $n$
iterations by $\{\boldsymbol{\Theta}^{(i)}\}_{i=1}^n$. The AM proposal at iteration $n+1$ is
\[
\boldsymbol{\Theta}' \sim \mathcal{N}\bigl(\boldsymbol{\Theta}^{(n)},\, s_d^2\Sigma_n + s_\epsilon^2 I_d\bigr),
\]
where $\Sigma_n$ is the sample covariance matrix (empirical covariance) of
$\{\boldsymbol{\Theta}^{(i)}\}_{i=1}^n$; $s_d^2$ is the scaling factor, commonly set to
$s_d^2 = (2.38)^2/d$ (an asymptotically optimal scaling for Gaussian targets;
\citealt{roberts1997weak}); and $s_\epsilon^2 I_d$ is a small regularization term (typically
$s_\epsilon^2=10^{-6}$ to $10^{-5}$) that ensures the proposal covariance is positive definite
and avoids collapse when $\Sigma_n$ is near-singular.

\paragraph{Algorithm (AM).}
\begin{enumerate}
  \item Initialize $\boldsymbol{\Theta}^{(0)}$ and set initial covariance $\Sigma_0 = \Sigma_{\text{init}}$
        (e.g., a diagonal matrix from prior scales or a small multiple of identity).
  \item For $n=0,1,2,\dots$:
    \begin{enumerate}
      \item Propose $\boldsymbol{\Theta}' \sim \mathcal{N}\bigl(\boldsymbol{\Theta}^{(n)},\,
            s_d^2\Sigma_n + s_\epsilon^2 I_d\bigr)$.
      \item Compute acceptance probability
      \[
      \alpha = \min\!\left(1,\; \frac{\pi(\boldsymbol{\Theta}')}{\pi(\boldsymbol{\Theta}^{(n)})}\right).
      \]
      \item With probability $\alpha$ set $\boldsymbol{\Theta}^{(n+1)}=\boldsymbol{\Theta}'$,
            else $\boldsymbol{\Theta}^{(n+1)}=\boldsymbol{\Theta}^{(n)}$.
      \item Update the sample mean and covariance $\Sigma_{n+1}$ using an online rank-one update
            ($O(d^2)$ per iteration). For numerical stability, regularize by adding $s_\epsilon^2 I_d$
            to the working covariance used in proposals.
    \end{enumerate}
\end{enumerate}

\subsection{Practical Modifications and Tuning}

Several practical modifications improve the behavior of the sampler. For the adaptive schedule,
adaptation begins only after a short pilot phase of $n_0$ iterations (e.g., $n_0=1{,}000$) to build
a reasonable initial covariance, and it may optionally be frozen after $N_\text{adapt}$ iterations to
produce final stationary draws. The scaling constant $s_d^2=(2.38)^2/d$ serves as a starting rule and
is reduced or increased if the empirical acceptance rate departs substantially from recommended
values (around $0.234$ for multivariate targets; \citealt{roberts1997weak,roberts2001optimal}). For
regularization, $s_\epsilon^2$ is set to a small value (e.g., $10^{-6}$ to $10^{-4}$) to avoid
degenerate covariance estimates, especially in early iterations or when parameters are highly
correlated. To respect diminishing adaptation and thereby satisfy ergodicity theory, the amount of adaptation is
reduced over time (e.g., by multiplying covariance updates by a factor $\gamma_n$ with
$\gamma_n\to0$ as $n\to\infty$) or adaptation is stopped after burn-in \citep{roberts2007coupling}. Finally, as a robust variant, the
Robust Adaptive Metropolis (RAM) algorithm of \citet{vihola2012robust} simultaneously adapts the
covariance shape and coerces the acceptance rate to a target, which can improve stability on
heavy-tailed or poorly scaled targets.

\subsection{Ergodicity and Theoretical Conditions}

Adaptive MCMC can preserve ergodicity (convergence to the correct posterior) when two key conditions
hold: (1) \emph{diminishing adaptation} (the magnitude of adaptation tends to zero), and (2)
\emph{containment} (the chain remains in regions where adaptation is controlled). Under such
conditions, the adaptive Metropolis algorithm is ergodic \citep{roberts2007coupling,roberts2009examples}.
In practice, we implement adaptation schedules (delayed start, decreasing adaptation gain, and
optional stop-after-burn-in) to respect these conditions.

\subsection{Implementation Details for the SEIARD Model}

Several implementation choices are specific to the SEIARD model. The chain is run on unconstrained
parameter transforms (log for positive parameters, logit for parameters in $(0,1)$) to ease
exploration and avoid boundary issues. The initial covariance $\Sigma_{\text{init}}$ is set as a
diagonal matrix whose elements are squared initial proposal widths (e.g., 0.1 to 1 times the
parameter scales on the transformed space). Multiple independent adaptive chains (e.g., four) are run
with different random seeds, with adaptation performed per chain; the chains are combined after
burn-in, and convergence is confirmed via the Gelman-Rubin diagnostic ($\widehat{R}$ close to 1).
Finally, for posterior prediction and forecasting, each retained posterior draw is used to simulate
the SEIARD model forward, producing predictive distributions for reported cases and deaths, from
which pointwise credible and prediction intervals are computed as empirical quantiles.

Adaptive Metropolis is simple to implement and works robustly for moderate-dimensional problems
(tens of parameters), particularly when derivatives of the likelihood with respect to parameters are
cumbersome or when compiled automatic-differentiation toolchains are not available.

\section{Results}
\label{sec:results}

\subsection{Implementation across U.S.\ States}
\label{sec:state_implementation}

We implemented the SEIARD model and Bayesian parameter estimation procedure separately for each of
the 50 U.S.\ states over the period from March to September 2020. For each state, daily time series
of confirmed COVID-19 cases, deaths, and testing rates were compiled from publicly available datasets
curated by the Johns Hopkins University Center for Systems Science and Engineering (JHU CSSE) and the
COVID Tracking Project \citep{johnhopkins2020dataset}. Population sizes were obtained from 2020 U.S.\
Census Bureau estimates to normalize incidence and testing rates.

To account for changes in population behavior and movement patterns, we incorporated state-level
Google Community Mobility Reports data, which quantify relative changes in visits to categories such
as retail and recreation, workplaces, parks, and residential areas. We constructed a composite
mobility index $M_t$ by averaging normalized mobility indicators (excluding residential categories to
avoid redundancy), scaled so that $M_t = 1$ corresponds to the baseline pre-pandemic mobility level
(February 2020). This index entered the SEIARD model through the time-dependent transmission rate
defined in Equation~(\ref{eq:beta_mobility}),
$\beta_t = \beta_0\,[1 + \alpha_M (M_t - 1)]\, e^{-\rho_\beta t}$, allowing the model to capture both
the influence of changing mobility and social activity on disease spread (through the
$\alpha_M(M_t-1)$ term) and gradual behavioral adaptation or residual control measures independent
of mobility, such as mask use (through the $e^{-\rho_\beta t}$ decay term). Because $\beta_t$
combines both effects, the fitted trajectory of $\beta(t)$ at the start of the series need not equal
the baseline parameter $\beta_0$ reported in Table~\ref{tab:seiard_parameters}: it reflects one
increment of the decay term together with the mobility deviation $M_t-1$ observed on that day, and
will only equal $\beta_0$ exactly if $M_t=1$ at $t=1$. This is visible, for example, in the New York
panel of Figure~\ref{fig:paramsNY}, where the plotted $\beta(t)$ at the start of the series differs
from the $\beta_0$ value in Table~\ref{tab:seiard_parameters} for this reason.

A few key epidemiological parameters were fixed based on established estimates from prior literature:
\[
\sigma = \frac{1}{5.5}, \quad
\gamma_I = \frac{1}{7}, \quad
\gamma_A = \frac{1}{5.5},
\]
where $\sigma$ denotes the rate at which exposed individuals become infectious, and $\gamma_I$ and
$\gamma_A$ represent the recovery rates for symptomatic and asymptomatic individuals, respectively.
These values were chosen in accordance with estimates reported in previous epidemiological studies on
COVID-19 transmission dynamics \citep{lauer2020incubation,li2020early,he2020temporal,byrne2020inferred}.

Each state model was fitted independently using the adaptive Metropolis algorithm described in
Section~\ref{sec:adaptivemcmc}. Posterior inference was performed on transformed parameter spaces
(log for positive parameters, logit for probabilities) to ensure numerical stability and
interpretability. For each state, four independent MCMC chains were initialized with overdispersed
starting points and run for 100{,}000 iterations, including 50{,}000 burn-in iterations discarded
prior to convergence assessment. Posterior draws were thinned every 10 iterations to reduce
autocorrelation. Adaptation of the proposal covariance was allowed during the first 20{,}000
iterations and subsequently frozen to satisfy the diminishing adaptation condition and ensure
ergodicity.

Convergence was monitored using the Gelman-Rubin diagnostic ($\widehat{R}$), effective sample
sizes (ESS), and trace plot inspection. Convergence was considered satisfactory when $\widehat{R}
< 1.1$ for all parameters and ESS exceeded 200. Computations were executed in parallel for all
50 states on a high-performance computing (HPC) cluster using one CPU core per state. The typical
runtime per state was approximately 1.5 to 2 hours on an Intel Xeon 2.6~GHz processor, depending on
the number of observations and numerical stiffness of the ODE solver. Posterior summaries (median
and 95\% credible intervals) were obtained from the retained samples, and posterior predictive
distributions were generated by forward simulation of the SEIARD system to produce 95\% prediction
intervals for daily cases and deaths.

For each state, we computed state-specific prediction intervals using the full Bayesian posterior
obtained from MCMC. Let $\{\boldsymbol{\Theta}^{(s)}\}_{s=1}^S$ denote $S$ posterior draws after
burn-in. For each draw $\boldsymbol{\Theta}^{(s)}$ we forward-simulated the deterministic SEIARD
system from the last observation day $T$ to forecast horizon $T+h$, producing model-predicted latent
trajectories $\widehat{C}^{(s)}_{t}$ and $\widehat{D}^{(s)}_{t}$ for $t=T+1,\dots,T+h$. To
incorporate observation uncertainty we generated observation-level replicates by drawing
\[
C^{\mathrm{pred},(s)}_{t}\sim\mathrm{NegBin}\bigl(\widehat{C}^{(s)}_{t},\kappa_C^{(s)}\bigr),\qquad
D^{\mathrm{pred},(s)}_{t}\sim\mathrm{NegBin}\bigl(\widehat{D}^{(s)}_{t},\kappa_D^{(s)}\bigr),
\]
using the overdispersion parameters $\kappa_C^{(s)},\kappa_D^{(s)}$ associated with
$\boldsymbol{\Theta}^{(s)}$. The empirical 2.5\% and 97.5\% percentiles of
$\{C^{\mathrm{pred},(s)}_{t}\}_{s=1}^S$ and $\{D^{\mathrm{pred},(s)}_{t}\}_{s=1}^S$ were taken as
the state-wise 95\% pointwise prediction intervals for daily reported cases and deaths, respectively.
All predictive summaries integrate parameter uncertainty (via the posterior draws) and observation
noise (via the Negative Binomial draws), providing state-specific measures of forecast uncertainty
\citep{gelman2013bayesian,carpenter2017stan}.

\subsection{State-Level Predictions and Time-Varying Parameters}

\begin{figure}[p]
    \centering
    \subfloat[Reported cases and deaths (with predicted mean and $95\%$ prediction interval).\label{fig:predAlabama}]{
        \includegraphics[width=0.9\textwidth]{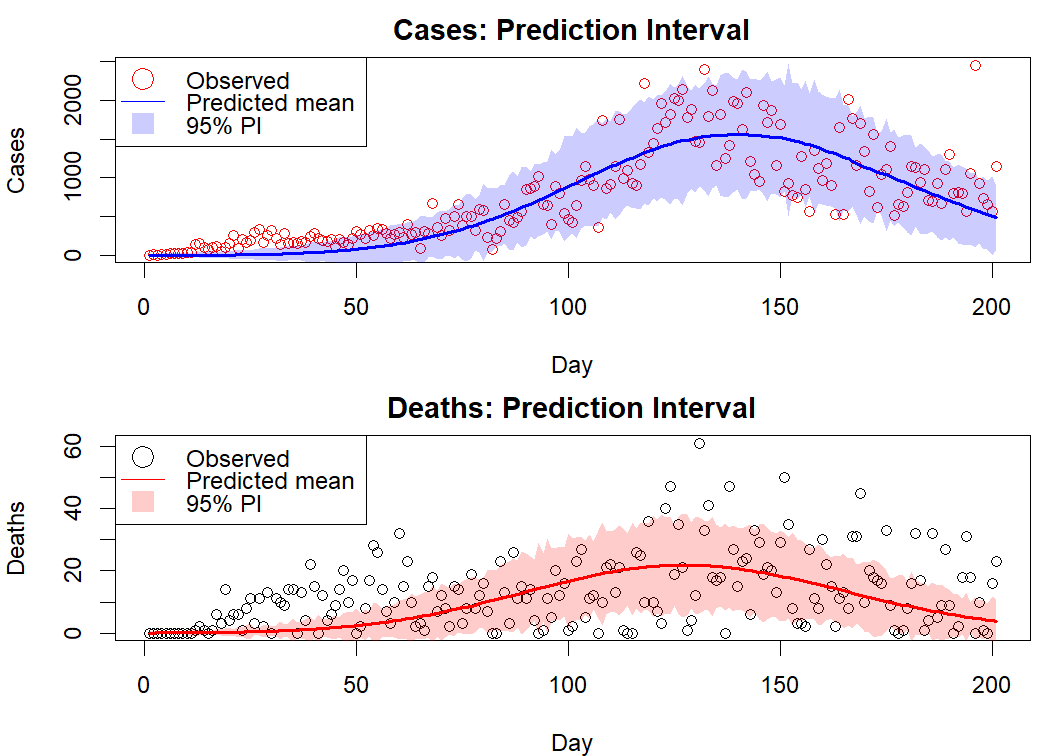}}\\[6pt]
    \subfloat[Time-varying parameters and effective reproduction rate
              ($\beta(t)$, $p(t)$, $\mu(t)$, and $R_e(t)$).\label{fig:paramsAlabama}]{
        \includegraphics[width=0.9\textwidth]{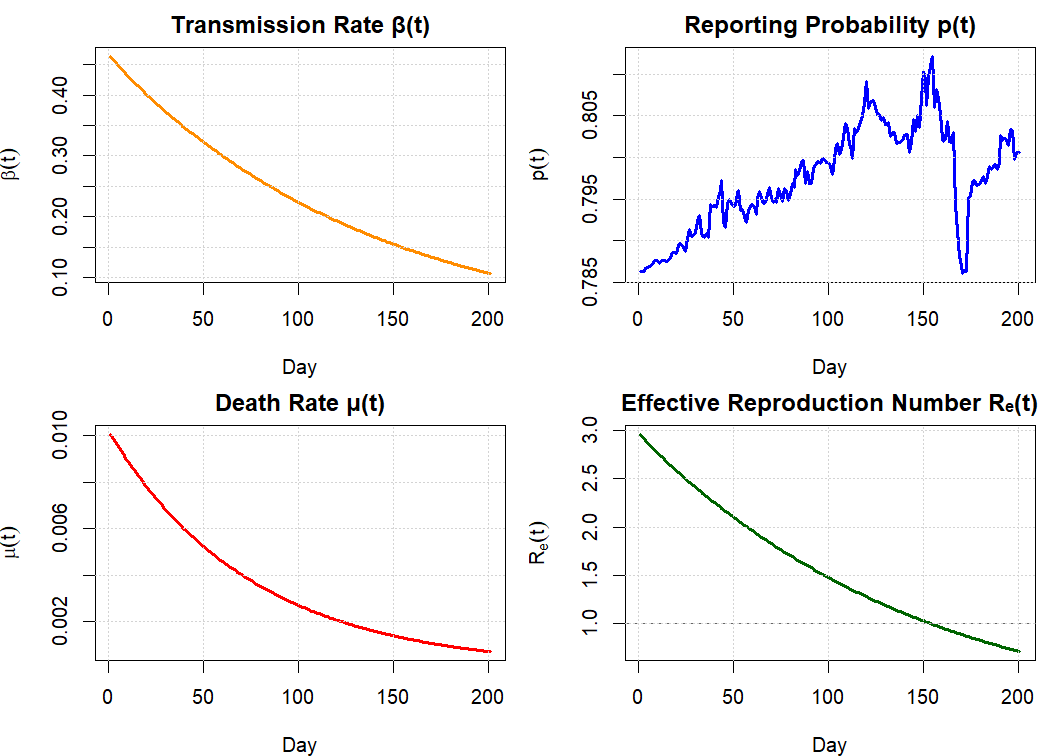}}
    \caption{Alabama: (\textbf{a}) Reported cases and predictions; (\textbf{b}) Time-varying
             parameters and $R_e(t)$.\label{fig:Alabama_all}}
\end{figure}

\begin{figure}[p]
    \centering
    \subfloat[Reported cases and deaths (with predicted mean and $95\%$ prediction interval).\label{fig:predCalifornia}]{
        \includegraphics[width=0.9\textwidth]{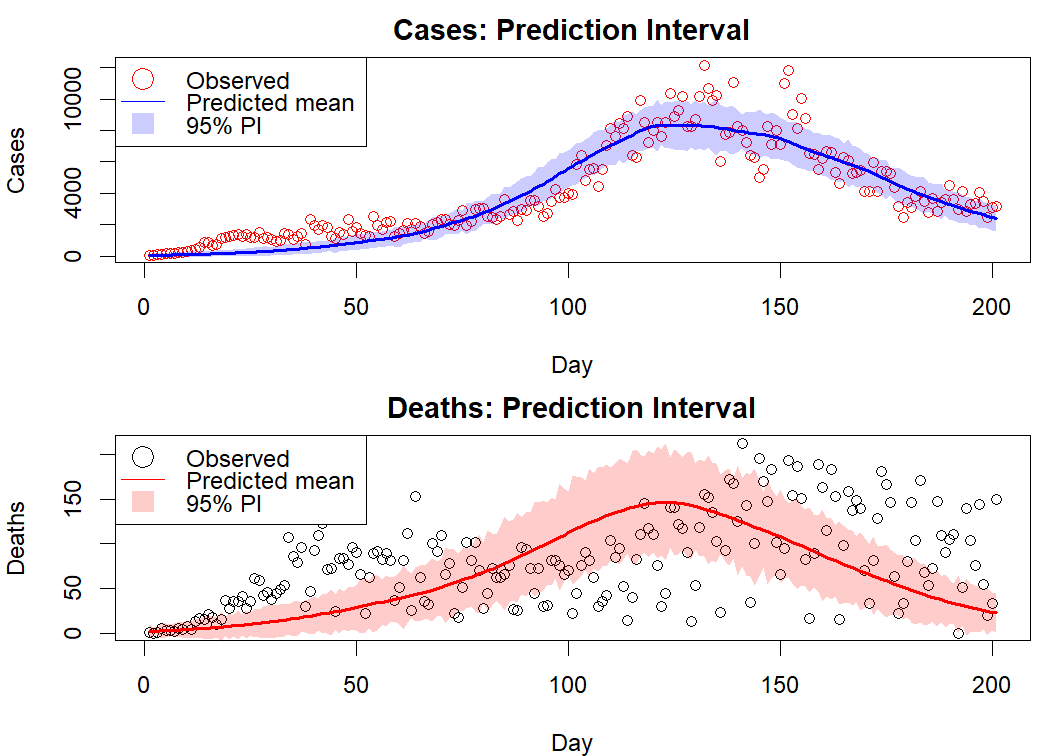}}\\[6pt]
    \subfloat[Time-varying parameters and effective reproduction rate
              ($\beta(t)$, $p(t)$, $\mu(t)$, and $R_e(t)$).\label{fig:paramsCalifornia}]{
        \includegraphics[width=0.9\textwidth]{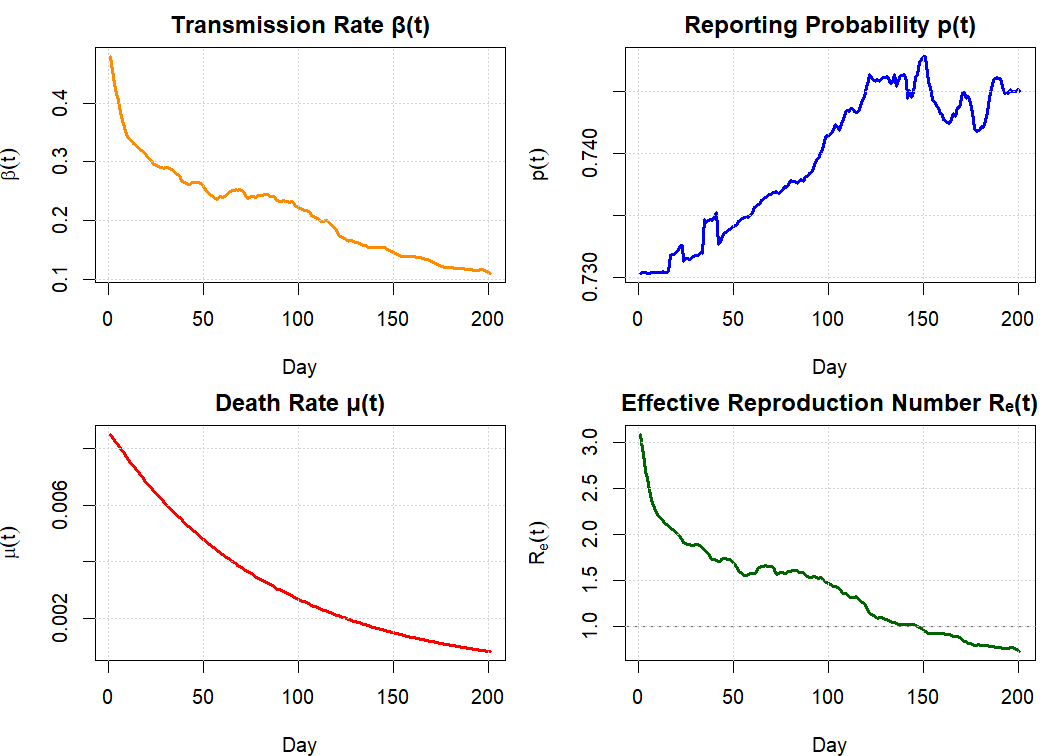}}
    \caption{California: (\textbf{a}) Reported cases and predictions; (\textbf{b}) Time-varying
             parameters and $R_e(t)$.\label{fig:California_all}}
\end{figure}

\begin{figure}[p]
    \centering
    \subfloat[Reported cases and deaths (with predicted mean and $95\%$ prediction interval).\label{fig:predNY}]{
        \includegraphics[width=0.9\textwidth]{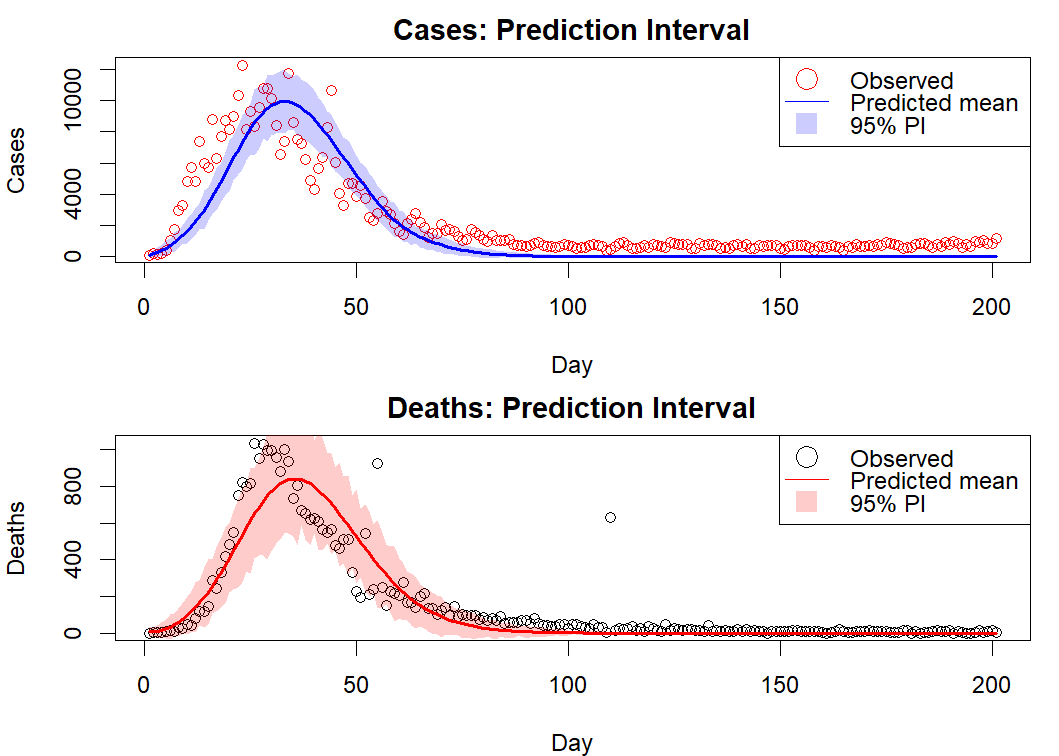}}\\[6pt]
    \subfloat[Time-varying parameters and effective reproduction rate
              ($\beta(t)$, $p(t)$, $\mu(t)$, and $R_e(t)$).\label{fig:paramsNY}]{
        \includegraphics[width=0.9\textwidth]{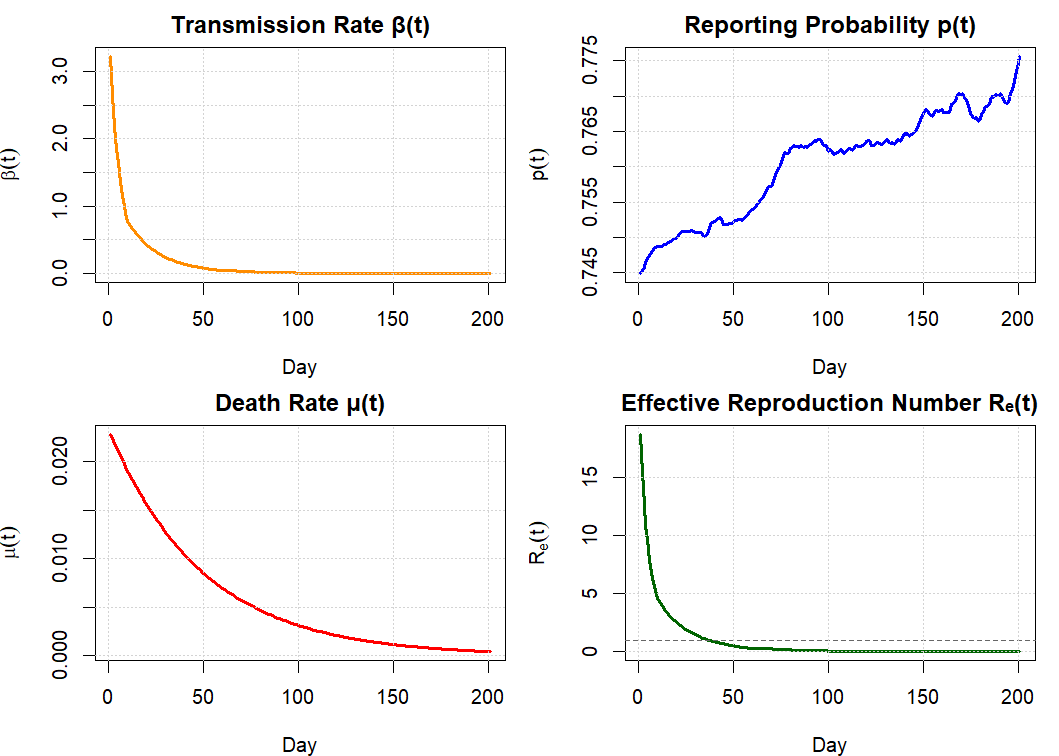}}
    \caption{New York: (\textbf{a}) Reported cases and predictions; (\textbf{b}) Time-varying
             parameters and $R_e(t)$.\label{fig:NY_all}}
\end{figure}

The posterior mean and 95\% prediction interval for daily infections and deaths are given in
Figures~\ref{fig:Alabama_all}, \ref{fig:California_all}, and \ref{fig:NY_all} for three
representative states: Alabama, California, and New York. The corresponding posterior means of the
time-varying coefficients are shown alongside in each figure.

\subsection{State-Wise Parameter Estimates}

Table~\ref{tab:seiard_parameters} presents the state-wise posterior medians of the fitted SEIARD
model parameters $\beta_0$, $\alpha_M$, $\rho_\beta$, $\mu_0$, and $\rho_\mu$, together with derived
epidemiological indicators, including the number of days the effective reproduction rate remained
above 1, its maximum value, total population, the Case Fatality Rate (CFR), the cumulative infection
rate, and the cumulative death rate, all derived from the posterior samples.

{\footnotesize
\setlength{\tabcolsep}{2pt}
\renewcommand{\arraystretch}{1.05}
\begin{longtable}{lrrrrrrrrrrr}
\caption{State-Wise SEIARD Model Fitted Parameters and Epidemiological Metrics. Population is
reported in millions.\label{tab:seiard_parameters}}\\
\toprule
State & $\beta_0$ & $\alpha_M$ & $\rho_{\beta}$ & $\mu_0$ & $\rho_{\mu}$
      & Days $R_e(t){>}1$ & Max $R_e(t)$ & Pop (M) & CFR (\%) & Inf.\ rate (\%) & Death (\%) \\
\midrule
\endfirsthead

\multicolumn{12}{c}{\tablename\ \thetable{} (continued from previous page)} \\
\toprule
State & $\beta_0$ & $\alpha_M$ & $\rho_{\beta}$ & $\mu_0$ & $\rho_{\mu}$
      & Days $R_e(t){>}1$ & Max $R_e(t)$ & Pop (M) & CFR (\%) & Inf.\ rate (\%) & Death (\%) \\
\midrule
\endhead

\midrule
\multicolumn{12}{r}{\textit{Continued on next page}} \\
\endfoot

\bottomrule
\endlastfoot

AL & 0.467 & 0.01  & 0.007 & 0.010 & 0.013 & 154 &  2.958 & 5.02  & 1.639 & 3.056 & 0.050 \\
AK & 0.488 & 0.353 & 0.006 & 0.014 & 0.020 & 171 &  2.505 & 0.73  & 0.613 & 1.156 & 0.007 \\
AZ & 1.371 & 0.750 & 0.013 & 0.003 & 0.001 & 120 &  6.514 & 7.15  & 2.586 & 3.042 & 0.079 \\
AR & 0.405 & 0.01  & 0.006 & 0.002 & 0.001 & 158 &  2.533 & 3.01  & 1.620 & 2.725 & 0.044 \\
CA & 0.580 & 0.416 & 0.007 & 0.009 & 0.012 & 146 &  3.084 & 39.54 & 1.920 & 2.061 & 0.040 \\
CO & 3.000 & 0.750 & 0.039 & 0.008 & 0.007 &  54 & 12.577 & 5.77  & 2.933 & 1.212 & 0.036 \\
CT & 5.000 & 0.729 & 0.050 & 0.024 & 0.011 &  51 & 19.156 & 3.61  & 7.880 & 1.585 & 0.125 \\
DC & 3.000 & 0.750 & 0.034 & 0.007 & 0.005 &  64 & 13.100 & 0.69  & 3.110 & 2.957 & 0.092 \\
DE & 3.000 & 0.750 & 0.026 & 0.020 & 0.018 &  69 & 11.264 & 0.99  & 4.088 & 1.542 & 0.063 \\
FL & 1.335 & 0.909 & 0.013 & 0.002 & 0.001 & 131 &  6.490 & 21.54 & 2.001 & 3.256 & 0.065 \\
GA & 0.410 & 0.01  & 0.006 & 0.009 & 0.010 & 150 &  2.493 & 10.71 & 2.276 & 2.791 & 0.064 \\
HI & 0.654 & 0.750 & 0.004 & 0.080 & 0.026 & 200 &  2.845 & 1.46  & 1.061 & 0.849 & 0.009 \\
ID & 0.704 & 0.592 & 0.009 & 0.009 & 0.014 & 152 &  3.710 & 1.84  & 1.113 & 2.276 & 0.025 \\
IL & 0.398 & 0.01  & 0.006 & 0.038 & 0.019 & 155 &  2.175 & 12.80 & 3.028 & 2.290 & 0.069 \\
IN & 1.200 & 0.750 & 0.015 & 0.080 & 0.026 & 123 &  4.555 & 6.79  & 2.978 & 1.777 & 0.053 \\
IA & 0.345 & 0.00  & 0.005 & 0.013 & 0.014 & 170 &  2.066 & 3.27  & 1.517 & 2.667 & 0.040 \\
KS & 0.383 & 0.01  & 0.005 & 0.013 & 0.016 & 161 &  2.291 & 2.94  & 1.072 & 2.035 & 0.022 \\
KY & 0.347 & 0.01  & 0.004 & 0.024 & 0.018 & 178 &  1.996 & 4.51  & 1.683 & 1.576 & 0.027 \\
LA & 2.500 & 0.750 & 0.043 & 0.080 & 0.036 &  56 & 10.154 & 4.66  & 3.295 & 3.571 & 0.118 \\
ME & 1.500 & 0.709 & 0.021 & 0.043 & 0.023 & 100 &  5.992 & 1.36  & 2.642 & 0.389 & 0.010 \\
MD & 3.000 & 0.750 & 0.026 & 0.040 & 0.024 &  88 & 10.733 & 6.18  & 3.166 & 2.013 & 0.064 \\
MA & 3.271 & 0.750 & 0.037 & 0.009 & 0.001 &  51 & 12.492 & 7.03  & 7.183 & 1.864 & 0.134 \\
MI & 4.847 & 0.750 & 0.045 & 0.073 & 0.029 &  55 & 16.151 & 10.08 & 5.194 & 1.347 & 0.070 \\
MN & 1.800 & 0.750 & 0.020 & 0.080 & 0.030 & 111 &  6.312 & 5.71  & 2.116 & 1.712 & 0.036 \\
MS & 0.553 & 0.01  & 0.008 & 0.017 & 0.012 & 146 &  3.258 & 2.96  & 3.010 & 3.277 & 0.099 \\
MO & 0.441 & 0.01  & 0.006 & 0.028 & 0.020 & 180 &  2.506 & 6.15  & 1.667 & 2.078 & 0.035 \\
MT & 0.363 & 0.457 & 0.004 & 0.008 & 0.010 & 201 &  1.945 & 1.08  & 1.380 & 1.183 & 0.016 \\
NE & 1.500 & 0.750 & 0.029 & 0.002 & 0.001 &  67 &  6.945 & 1.96  & 1.092 & 2.273 & 0.025 \\
NV & 1.500 & 0.750 & 0.013 & 0.080 & 0.029 & 122 &  5.810 & 3.10  & 2.000 & 2.555 & 0.051 \\
NH & 1.500 & 0.712 & 0.022 & 0.015 & 0.007 &  83 &  6.487 & 1.38  & 5.348 & 0.596 & 0.032 \\
NJ & 5.500 & 0.696 & 0.054 & 0.010 & 0.001 &  45 & 20.702 & 9.29  & 7.823 & 2.217 & 0.173 \\
NM & 1.500 & 0.750 & 0.019 & 0.080 & 0.028 & 106 &  6.001 & 2.12  & 3.012 & 1.369 & 0.041 \\
NY & 5.773 & 0.750 & 0.061 & 0.022 & 0.019 &  35 & 19.697 & 20.20 & 7.099 & 2.283 & 0.162 \\
NC & 0.437 & 0.00  & 0.007 & 0.006 & 0.010 & 144 &  2.686 & 10.44 & 1.664 & 1.997 & 0.033 \\
ND & 0.338 & 0.01  & 0.003 & 0.021 & 0.017 & 201 &  1.967 & 0.78  & 1.129 & 2.694 & 0.030 \\
OH & 0.486 & 0.01  & 0.008 & 0.034 & 0.019 & 147 &  2.702 & 11.80 & 3.126 & 1.287 & 0.040 \\
OK & 0.424 & 0.01  & 0.006 & 0.013 & 0.017 & 176 &  2.537 & 3.96  & 1.182 & 2.152 & 0.025 \\
OR & 0.419 & 0.349 & 0.006 & 0.011 & 0.014 & 152 &  2.208 & 4.24  & 1.664 & 0.779 & 0.013 \\
PA & 5.000 & 0.750 & 0.048 & 0.009 & 0.002 &  52 & 20.177 & 13.00 & 5.061 & 1.243 & 0.063 \\
RI & 3.764 & 0.750 & 0.041 & 0.007 & 0.001 &  53 & 15.464 & 1.10  & 4.545 & 2.226 & 0.101 \\
SC & 0.485 & 0.238 & 0.007 & 0.004 & 0.004 & 145 &  2.908 & 5.12  & 2.279 & 2.861 & 0.065 \\
SD & 0.300 & 0.01  & 0.003 & 0.080 & 0.026 & 201 &  1.528 & 0.89  & 1.003 & 2.452 & 0.025 \\
TN & 0.393 & 0.00  & 0.006 & 0.002 & 0.002 & 157 &  2.465 & 6.91  & 1.237 & 2.764 & 0.034 \\
TX & 0.446 & 0.01  & 0.007 & 0.002 & 0.001 & 154 &  2.926 & 29.15 & 2.055 & 2.655 & 0.055 \\
UT & 0.399 & 0.01  & 0.006 & 0.006 & 0.014 & 153 &  2.460 & 3.21  & 0.634 & 2.228 & 0.014 \\
VT & 2.500 & 0.750 & 0.033 & 0.080 & 0.040 &  65 &  9.359 & 0.64  & 3.324 & 0.271 & 0.009 \\
VA & 1.800 & 0.750 & 0.020 & 0.080 & 0.035 & 105 &  6.505 & 8.63  & 2.164 & 1.698 & 0.037 \\
WA & 0.483 & 0.538 & 0.006 & 0.017 & 0.017 & 136 &  2.163 & 7.71  & 2.427 & 1.176 & 0.029 \\
WV & 0.399 & 0.00  & 0.006 & 0.018 & 0.013 & 164 &  2.344 & 1.79  & 2.173 & 0.865 & 0.019 \\
WI & 0.372 & 0.01  & 0.005 & 0.014 & 0.018 & 182 &  2.222 & 5.89  & 1.040 & 2.109 & 0.022 \\
WY & 0.327 & 0.01  & 0.004 & 0.011 & 0.016 & 182 &  2.222 & 0.58  & 0.869 & 0.997 & 0.009 \\
\end{longtable}
}

Table~\ref{tab:seiard_summary} presents descriptive statistics for the fitted SEIARD model
parameters across the 50 U.S.\ states.

\begin{table}[H]
\centering
\caption{Descriptive Summary of SEIARD Model Parameters Across U.S.\ States.\label{tab:seiard_summary}}
\begin{tabular}{lrrrr}
\toprule
\textbf{Parameter} & \textbf{Mean} & \textbf{Std.\ Dev.} & \textbf{Min} & \textbf{Max} \\
\midrule
$\beta_0$ (Baseline transmission rate)         & 1.518  & 1.558  & 0.300 &  5.773 \\
$\alpha_M$ (Mobility effect)                   & 0.411  & 0.353  & 0.000 &  0.909 \\
$\rho_{\beta}$ (Rate of change in $\beta$)     & 0.017  & 0.016  & 0.003 &  0.061 \\
$\mu_0$ (Baseline mortality rate)              & 0.0268 & 0.0279 & 0.002 &  0.080 \\
$\rho_{\mu}$ (Rate of change in $\mu$)         & 0.0150 & 0.0104 & 0.001 &  0.040 \\
Days $R_e(t)>1$                                & 125.92 & 49.78  & 35.0  & 201.0  \\
Max $R_e(t)$                                   & 6.320  & 5.540  & 1.528 & 20.702 \\
CFR (\%)                                       & 2.681  & 1.824  & 0.613 &  7.880 \\
Infection rate (\%)                            & 1.961  & 0.795  & 0.271 &  3.571 \\
Death rate (\%)                                & 0.051  & 0.039  & 0.007 &  0.173 \\
\bottomrule
\end{tabular}
\end{table}

The baseline transmission rate ($\beta_{0}$) varied substantially among states, ranging from
approximately 0.3 to 5.8. States such as New York, Pennsylvania, and New Jersey exhibited
relatively higher $\beta_{0}$ values, indicating rapid infection spread during the initial outbreak
phases. In contrast, regions such as Alabama, Arkansas, and Alaska showed lower transmission
potential, reflecting smaller population densities or stronger early containment measures.

The mobility-related adjustment factor ($\alpha_{M}$) quantifies the sensitivity of transmission
to behavioral and mobility variations. States with higher $\alpha_{M}$ values experienced stronger
correlations between changes in mobility and infection rate, suggesting that social distancing and
movement restrictions played a significant role in mitigating spread.

The temporal change in transmission ($\rho_{\beta}$) captures the reduction in infection potential
over time, likely associated with public health interventions and behavioral adaptation. The initial
mortality parameter ($\mu_{0}$) was generally low, indicating that baseline death probabilities
were modest in most states. The mortality change rate ($\rho_{\mu}$) remained positive but small,
suggesting gradual temporal improvement in disease outcomes.

The maximum effective reproduction number ($R_{e}^{\text{max}}$) ranged between approximately 1.5
and 20.7 across states. This variation highlights significant differences in local epidemic
potential. High $R_{e}^{\text{max}}$ values, observed in several northeastern states (e.g., New
York, New Jersey, Pennsylvania), imply that uncontrolled spread led to rapid case doubling, while
states with lower $R_{e}^{\text{max}}$ likely benefited from early interventions.

In summary, baseline transmission ($\beta_{0}$) varied widely across states, revealing substantial
heterogeneity in transmission potential; initial mortality ($\mu_{0}$) was low in most states,
suggesting effective early containment or healthcare responses; and peak effective reproduction
numbers ($R_{e}^{\text{max}}$) ranged from 1.5 to 20.7, indicating that uncontrolled spread could
double infections every few days in high-transmission regions. As shown in
Table~\ref{tab:seiard_parameters}, larger states such as California and Texas combined moderate
$\beta_{0}$ with a high total infection burden owing to population size, while infection rates
typically fell between 1\% and 3\% and CFR values remained under 8\%, broadly aligning with
national-level observations.

Overall, the model parameters reveal that variations in population density, behavioral responses,
and healthcare access collectively shaped the observed infection and mortality trajectories. These
results can inform future preparedness strategies by highlighting the need for flexible control
measures responsive to regional dynamics.

\subsection{Clustering of States by Epidemiological Characteristics}
\label{sec:clustering}

To investigate similarities and heterogeneity in epidemic dynamics across U.S.\ states, we
performed a clustering analysis based on the posterior estimates of key epidemiological parameters.
For each state, we summarized the posterior distributions of the baseline transmission rate
($\beta_0$), the maximum effective reproduction number over the study period (${R}_{et}^{max}$), the number of Days effective reproduction number is greater than 1 ($Rg1$), the mobility effect ($\alpha_M)$, Baseline mortality rate ($\mu_0$). 
Rate of change in $\mu$ ($\rho_{\mu}$), infection rate,  death rate, the case fatality rate (CFR), computed as the ratio of cumulative deaths to cumulative infections inferred from the latent model compartments.

Posterior medians of the selected parameters were first standardized (zero mean and unit variance)
to remove scale differences. We then computed pairwise Euclidean distances in the standardized
parameter space and applied hierarchical agglomerative clustering using Ward's minimum-variance
linkage criterion. The optimal number of clusters was selected using a combination of the silhouette
coefficient and the gap statistic. Clusters were visualized via two-dimensional projections from
principal component analysis (PCA) on the standardized parameter matrix.

\begin{figure}[H]
    \centering
    \includegraphics[width=0.9\textwidth]{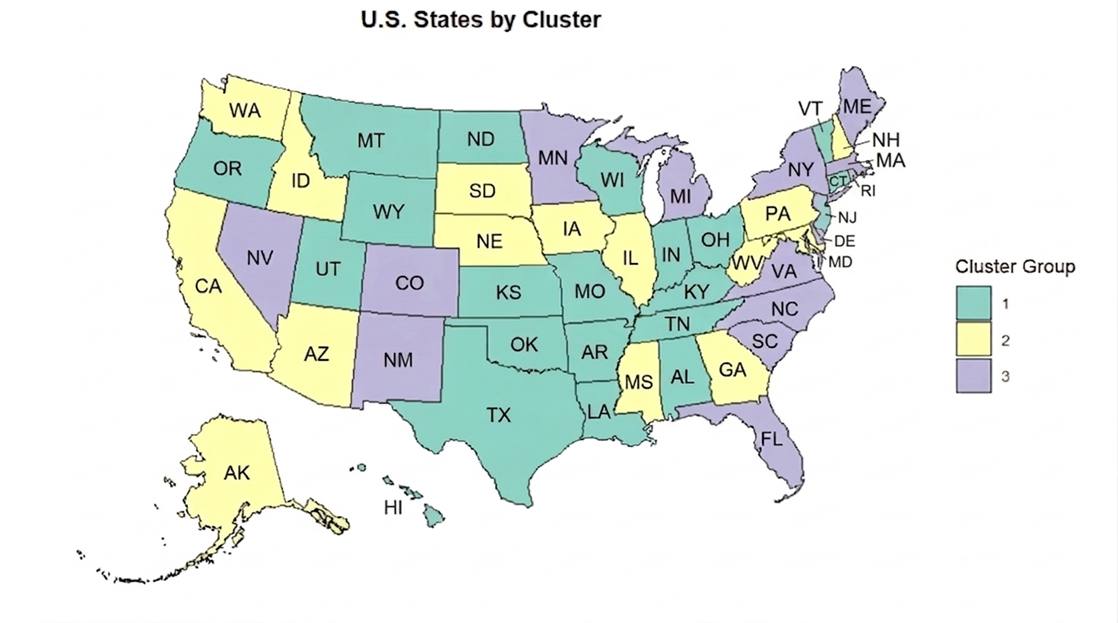}
    \caption{Cluster of U.S.\ states based on epidemiological parameters.\label{fig:uscluster1}}
\end{figure}

\begin{figure}[H]
    \centering
    \includegraphics[width=0.9\textwidth]{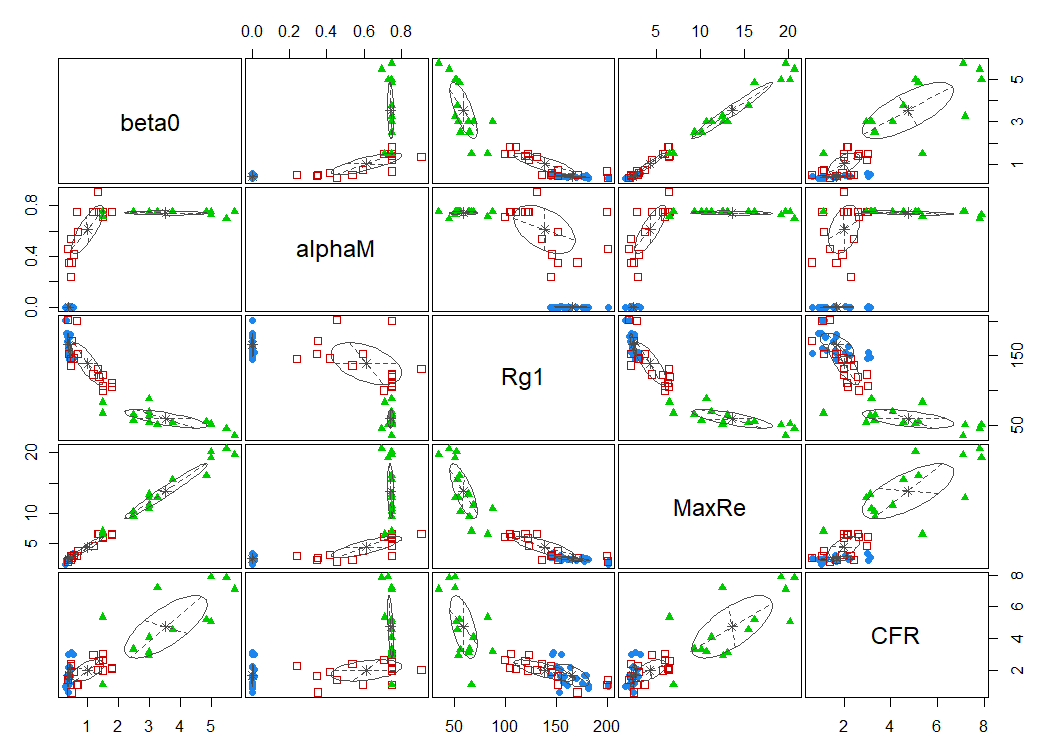}
    \caption{Robust cluster groups.\label{fig:uscluster2}}
\end{figure}

A three-cluster classification was obtained from the fitted SEIARD parameters and epidemiological
indicators, as illustrated in Figures~\ref{fig:uscluster1} and \ref{fig:uscluster2}.

Cluster~1 (teal) includes a large portion of the central and southeastern states, such as Texas,
Oklahoma, Missouri, and Wisconsin. States exemplifying this cluster generally exhibited $\beta_{0}$
values below 0.5 and comparatively stable reproduction numbers ($R_{e}^{\text{max}}$ typically
below 4); the mortality parameters were also moderate, suggesting balanced control between infection
spread and fatality outcomes. These states reflect regions where interventions mitigated severe
outbreaks despite ongoing transmission.

Cluster~2 (yellow) comprises many northern and western states, such as California, Washington, South Dakota,
and Alaska, and displayed lower infection and mortality levels. These states had relatively small
$\beta_{0}$ values and lower $R_{e}^{\text{max}}$, indicating controlled transmission dynamics and
slower epidemic growth; early policy actions, higher compliance with mobility restrictions, and lower
population density may have contributed to this profile.

Cluster~3 (lavender) primarily included several coastal and high-density states such as New York,
Virginia, and parts of the western and southern U.S. These states were characterized by higher
transmission and mortality parameters ($\beta_{0} > 1.0$, elevated $\mu_{0}$, and $R_{e}^{\text{max}}$
often exceeding 5). The combination of dense urban populations and initial outbreak severity placed
these states in a high-impact category, emphasizing the challenges of containing community spread
once transmission had intensified.

Overall, the cluster analysis highlights substantial spatial heterogeneity in epidemic behavior
across the United States. The results underscore that while many states maintained moderate
transmission and fatality levels, others faced accelerated infection dynamics due to demographic
and mobility factors. These distinctions justify the use of region-specific control strategies
rather than uniform nationwide policies.

\section{Discussion}
\label{sec:discussion}

The proposed SEIARD model provides a flexible framework for representing how public health
interventions, changing testing intensity, and evolving clinical care collectively influence
observed epidemic curves. By coupling time-varying transmission, reporting, and mortality rates to
observable covariates, namely mobility and testing, the model moves beyond static parameterizations
and tracks the dynamic interplay between human behavior and disease spread. The model effectively
captured the regional heterogeneity of the COVID-19 epidemic across U.S.\ states. Wide variation
in the baseline transmission rate ($\beta_{0}$), mortality rate ($\mu_{0}$), and maximum effective
reproduction number ($R_{e}^{\text{max}}$) highlights the influence of demographic structure,
healthcare capacity, and intervention intensity at the state level, underscoring the importance
of context-specific, adaptive public health responses rather than uniform nationwide strategies.

The Bayesian approach adopted here offers several advantages over deterministic
calibration methods. Parameter uncertainty is propagated through to all derived quantities,
including $R_e(t)$ and the prediction intervals for daily cases and deaths, enabling
more honest communication of forecast uncertainty to decision-makers. The use of adaptive
Metropolis MCMC made inference tractable without requiring gradient information or
automatic-differentiation toolchains, which is an important practical consideration when the
likelihood surface is defined implicitly through an ODE solver.

The state-level clustering analysis revealed three distinct epidemiological profiles across the
United States. Northeastern states characterized by high population density and early epidemic
onset formed a high-transmission, high-impact cluster, while many central and western states
exhibited more moderate and controlled transmission dynamics. These groupings align with the
variation in $\beta_0$ and $R_{e}^{\text{max}}$ estimated by the SEIARD model, providing convergent
evidence that geographic and demographic factors are primary drivers of epidemic heterogeneity.
The clustering methodology proved effective at separating these epidemiological profiles despite
the considerable variation in data quality across state surveillance systems.

Taken together, the mechanistic SEIARD modeling and the data-driven clustering analysis offer
complementary perspectives on COVID-19 dynamics across U.S.\ states. The model provides a
process-level understanding of transmission and intervention effects, while the clustering reveals
the regional structure underlying epidemic severity. Both analyses point to the same policy
implication: effective pandemic preparedness requires regionally tailored strategies that account
for demographic composition, healthcare infrastructure, and behavioral responsiveness to public
health guidance.

\subsection{Limitations and further model extensions}
A limitation of the approach is
the assumption of homogeneous mixing within each state; incorporating age-structured contact
matrices or spatial metapopulation structure would likely improve fit in heterogeneous
urban and rural settings. Additional extensions that would strengthen the framework include explicit
hospitalization compartments, reporting delay distributions, and vaccination compartments covering
the post-2020 period. In particular, the covariate framework can be generalized further by allowing $\beta(t)$ and
$p(t)$ to depend on richer predictor sets through log-linear or logistic regression submodels,
\[
\log \beta(t) = \beta_0 + \sum_j \alpha_j X_j(t),
\qquad
\operatorname{logit} p(t) = p_0 + \sum_k \eta_k Z_k(t),
\]
where $X_j(t)$ and $Z_k(t)$ represent additional time-varying covariates such as policy stringency
indices, vaccination coverage, or meteorological variables. This general structure nests the
specific parameterizations in Equations~(\ref{eq:beta_mobility}) and~(\ref{eq:report_testing}) as
special cases, offering a pathway to systematically incorporate emerging data streams into the model.

\section{Conclusions}
\label{sec:conclusions}

This paper presented a unified framework for retrospective analysis of COVID-19 dynamics across
U.S.\ states, integrating mechanistic compartmental modeling, Bayesian inference, and clustering
of state-level epidemiological profiles.

The extended SEIARD model with time-varying transmission, reporting, and mortality rates
successfully captured state-level epidemic trajectories over the period March to September 2020.
Substantial heterogeneity in estimated parameters across states confirms that a single national
model would obscure important regional variation in transmission potential and intervention
effectiveness. The Bayesian adaptive MCMC framework provided full posterior distributions for all
parameters, yielding calibrated uncertainty quantification for both in-sample fits and predictive
intervals, a meaningful improvement over point-estimate calibration methods used in much of the
prior literature.

The clustering analysis identified three distinct epidemiological profiles among U.S.\ states,
grouping them by transmission intensity, reproduction number, and mortality burden. These clusters
correspond broadly to geographic and demographic characteristics, suggesting that future
preparedness planning could benefit from tailoring response strategies to regional cluster
membership rather than applying uniform federal policies.

Taken together, these results demonstrate that integrating mechanistic transmission modeling with
state-level Bayesian inference and clustering yields a coherent, regionally resolved picture of the
early COVID-19 epidemic in the United States. The framework is readily extensible to additional
covariates, vaccination dynamics, and later pandemic phases, and it offers a practical basis for
demographic-sensitive, regionally tailored preparedness planning for future public health
emergencies.

\vspace{6pt}

\authorcontributions{Conceptualization, P.B.; methodology, P.B.; software, P.B.; formal analysis,
P.B.; investigation, P.B.; data curation, P.B.; writing, original draft preparation, P.B.;
writing, review and editing, P.B.; visualization, P.B.; funding acquisition, P.B. The author has
read and agreed to the published version of the manuscript.}

\funding{No external funding was available to support this research}

\institutionalreview{Not applicable.}

\informedconsent{Not applicable.}

\dataavailability{Data used in this study are publicly available from the Johns Hopkins University
CSSE COVID-19 repository (\url{https://github.com/CSSEGISandData/COVID-19}).}


\acknowledgments{The author gratefully acknowledges support from the Grauel Fellowship at John Carroll University which supported this project by providing a teaching release to focus exclusively on research. }

\conflictsofinterest{The author declares no conflicts of interest.}

\abbreviations{Abbreviations}{
The following abbreviations are used in this manuscript:\\

\noindent
\begin{tabular}{@{}ll}
SEIARD     & Susceptible-Exposed-Infectious-Asymptomatic-Recovered-Deceased \\
COVID-19   & Coronavirus Disease 2019 \\
SARS-CoV-2 & Severe Acute Respiratory Syndrome Coronavirus 2 \\
SEIR       & Susceptible-Exposed-Infectious-Recovered \\
MCMC       & Markov Chain Monte Carlo \\
ODE        & Ordinary Differential Equation \\
AM         & Adaptive Metropolis \\
RAM        & Robust Adaptive Metropolis \\
JHU CSSE   & Johns Hopkins University Center for Systems Science and Engineering \\
ESS        & Effective Sample Size \\
HPC        & High-Performance Computing \\
CFR        & Case Fatality Rate \\
PCA        & Principal Component Analysis \\
\end{tabular}
}

\begin{adjustwidth}{-\extralength}{0cm}

\reftitle{References}

\bibliographystyle{unsrtnat}
\bibliography{references}

\end{adjustwidth}
\end{document}